\documentclass[twocolumn]{aastex7}

\usepackage{makecell}
\usepackage{comment}
\usepackage{natbib}
\usepackage{multirow}
\usepackage{booktabs}
\usepackage{epsfig}
\usepackage{ulem}
\usepackage{rotating}
\usepackage{graphicx}
\usepackage{url}
\usepackage{xcolor}
\usepackage{amsmath}
\usepackage{color}
\usepackage{savesym}
\savesymbol{tablenum}
\usepackage{siunitx}
\usepackage[normalem]{ulem}
\usepackage{appendix}
\usepackage{enumerate}
\usepackage{subcaption}
\usepackage{gensymb}
\usepackage{xspace}
\usepackage{longtable}
\usepackage{afterpage}
\usepackage{threeparttable}
\usepackage{tabularx}
\usepackage[bottom]{footmisc}

\defcitealias{Ginolin_2025_ZTFcolor}{G25a}
\defcitealias{Ginolin_2025_ZTFstretch}{G25b}
\defcitealias{BroutScolnic21}{BS21}
\newcommand{\GinolinColor}{\citetalias{Ginolin_2025_ZTFcolor}\xspace}
\newcommand{\GinolinStretch}{\citetalias{Ginolin_2025_ZTFstretch}\xspace}
\newcommand{\BSdust}{\citetalias{BroutScolnic21}\xspace}

\newcommand\logmass{\log_{10}(\frac{M*}{M_\odot})}
\newcommand\lcdm{\Lambda\text{CDM}}

\DeclareSIUnit\parsec{pc}
\DeclareSIUnit\h{\textit{h}}

\graphicspath{{./}{fig/}}

\begin{document}

    \title{Reassessing the ZTF Volume-Limited Type Ia Supernova Sample and Its Implications for \\Continuous, Dust-Dependent Models of Intrinsic Scatter}

    \author[0000-0002-6124-1196]{Yukei S. Murakami}
    \affiliation{Department of Physics and Astronomy, Johns Hopkins University, Baltimore, MD 21218, USA}
    \email{ymuraka2@jhu.edu}
    
    \author[0000-0002-4934-5849]{Daniel Scolnic}
    \affiliation{Department of Physics, Duke University, Durham, NC 27708, USA}
    \email{daniel.scolnic@duke.edu}

\begin{abstract}
The second Data Release from Zwicky Transient Facility includes light curves of 3628 Type Ia supernovae (SNe~Ia), making it the largest low-redshift (mostly $z\lesssim0.15$) SNe Ia sample available. One central question for analyses of SNe Ia is whether the remaining diversity of standardized luminosities arises in part from an intrinsic or extrinsic effect --- characterized by the color-independent bimodality in progenitor population or color- and host- dependent diversity in dust extinction, respectively. 
In the initial analyses of the volume-limited subset ($z<0.06$; 945 SNe) of this sample from the ZTF collaboration, the authors reported evidences for the former hypothesis, in contrast to many of the previous evidences for the dust hypothesis found across other largest high- and low- redshift SNe~Ia samples.
We re-analyze the volume-limited ZTF SNe~Ia in the same manner that previous samples were analyzed and report consistency with trends seen in literature samples and in support of the dust hypothesis. 
We find the following:
1. a color dependency in the canonical `mass-step' size for SNe, with red SNe having a larger host-dependent residual step than blue SNe by $0.18\pm0.09$ mag; 
2. a color-dependent difference in the Hubble residual scatter, with red SNe having a $\sim33\%$ larger scatter than blue SNe at $>3\sigma$ significance;
3. data's preference of a model that accounts for color-dependency over a simple `step' model at $\sim3\sigma$;
4. the strongest evidence to date ($3.5\sigma$) that the relationship between SN color, host-galaxy properties, and luminosity is continuous rather than characterized by a discrete step. Accounting for 3 and 4 with our new model, Host2D, yields a $4.0\sigma$ improvement over the mass-step model.
We trace the difference in reported findings to the fitting and analysis methods, in particular the model complexity allowed for the color-luminosity relation, rather than a difference in the sample itself. 

\end{abstract}

\section{Introduction}

Type Ia supernovae (SNe~Ia), thermonuclear explosion of white dwarfs, are arguably one of the most successful distance indicators in the observational cosmology \citep[see, e.g.,][for review]{Filippenko_2005_SNIacosmology,Branch_Wheeler_2017_Supernova,Jha_2019_SNIa_review}.
A large fraction of its intrinsic variation in luminosities ($\sim0.4$~mag) can be removed by the Tripp standardization \citep{Tripp99}, a linear subtraction of stretch--luminosity and color--luminosity relations, and the small size of the resulting scatter ($\sim0.15$~mag) has contributed to the discovery of accelerating expansion in the universe and subsequent development of the observational cosmology \citep[e.g.,][]{Riess_1998_DE,Perlmutter_1999_DE,Betoule_2014_JLA,Brout22}. 
In the era of precision cosmology with numerous tensions and possible implications for physics beyond $\Lambda$CDM \citep[e.g.,][]{Riess_2022_SH0ESmain,Murakami_2023_SIP, Breuval_2024_SMC,DESY5_2024,DESI_DR2_cosmo_2025}, it is critical to understand the origin of this remaining $\sim0.15$~mag scatter to further reduce it in a quest for improved cosmological constraints.

\begin{figure}
    \centering
    \includegraphics[width=\linewidth]{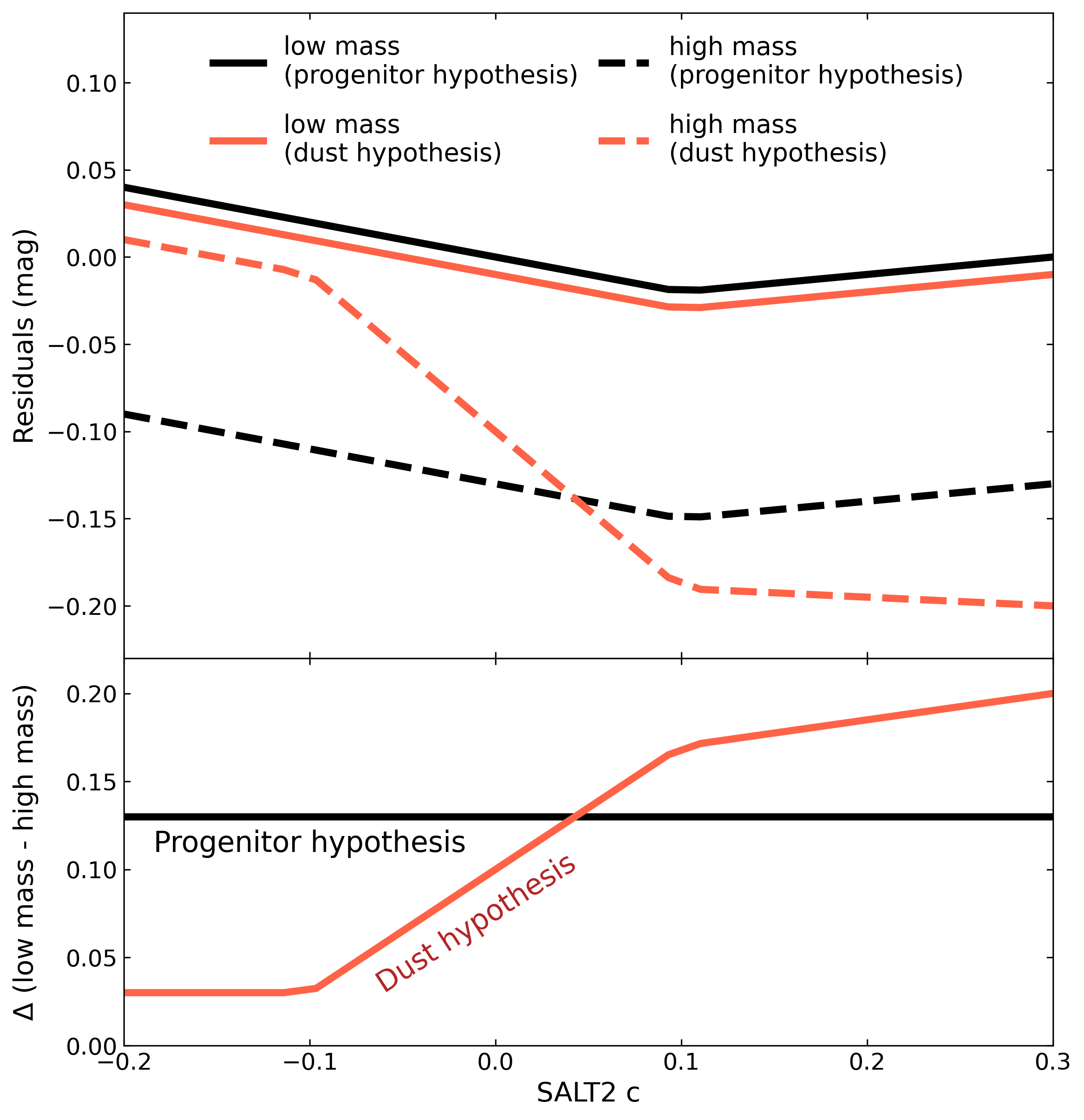}
    \caption{A conceptual illustration of the trends in residuals that the dust and progenitor hypotheses predict, over a range of SN~Ia color. \textit{Top panel:} The predicted residuals of SNe~Ia, split by their host property (e.g., stellar mass). \textit{Bottom panel:} The predicted separation between the split populations (i.e., the separation between solid curve and the dashed curve in the top panel). Progenitor hypothesis predicts a color-independent, constant offset, while the dust hypothesis predicts a color-dependent increase of the separation.}
    \label{fig:hypotheses_illustration}
\end{figure}

Significant relations between this leftover residuals and the properties of galaxies that host each SN have been reported (originally found in studies like \citealp{Kelly2010, Lampeitl2011, Sullivan2010} but also many other, e.g., \citealp{Childress_2013_massstep,Uddin_2020_massslope,Murakami_2021_host,Zhang_Murakami_2021_slopefit}), and there are two proposed hypotheses that describe the origin of this scatter.
The first hypothesis (comprehensively presented in \citealp{Rigault13}) states that the host galaxy properties are linked to different progenitor channels of SNe Ia, and a `step'--like drop in the mean Hubble residual versus host-galaxy properties indicates that there are two separate progenitor channels.  In this model, any further residual scatter after accounting for this effect is `gray' luminosity variation, independent of color (Fig.1, black curves).
In the second hypothesis (comprehensively presented in \citealp{BroutScolnic21}; hereafter \BSdust), the dependence of Hubble residuals on host-galaxy properties is driven by different dust properties in different host galaxies. In this model, the scatter is mainly due to differences in the reddening properties (commonly parameterized by the slope of optical extinction law, $R_V$), of the dust that interact with the light from SNe~Ia. 
This predicts an ``alligator-mouth'' profile (Fig.~\ref{fig:hypotheses_illustration}, red curves), where there is little difference in Hubble residuals between SNe with low-mass and high-mass hosts for bluer SNe due to negligible dust extinction, followed by a large divergence in the Hubble residuals toward redder SNe as larger reddening makes them more sensitive to the variation in extinction laws.  
Evidence supporting the latter hypothesis have been found in multiple analyses of large SNe~Ia samples (e.g. \citealp{BroutScolnic21,Rubin23}; see Fig. 6 of \BSdust, Fig. 2 of \citealp{Popovic23}, and Fig. 5 of \citealp{Vincenzi24} for the reported alligator-mouth profiles).

The Zwicky Transient Facility (ZTF, \citealp{Bellm19}) provides an excellent opportunity to test these two hypotheses. 
Recently, the ZTF Cosmology Science Working Group released a compilation of light curves in a second data release (DR2, \citealp{Rigault25}). Of the total 3628 SNe~Ia included in DR2, about $\sim2600$ SNe are expected to be useful for cosmology after the final calibration in DR2.5, which is to be released in the near future.
Within this sample, on the order of $\sim1000$ SNe lie within the volume limit range of $z<0.06$.  As this volume-limited subsample provides the largest data set of SNe~Ia that is not strongly affected by selection biases, the ZTF collaboration has also provided a first-round of analyses of various insights into this sample, including comparisons of trends of supernova properties previously seen in the literature \citep[e.g.,][]{Ginolin_2025_ZTFcolor,Ginolin_2025_ZTFstretch}.  
One area that has been particularly highlighted by the ZTF collaboration is the results presented in \cite{Ginolin_2025_ZTFcolor} (hereafter \GinolinColor), in which there are claims that the previously found evidences for the dust hypothesis is insignificant.

In \GinolinColor, the authors perform a variety of diagnostics with the volume-limited ZTF SNe to compare these two theories. For instance, they find that the color-luminosity slope, referred to as the $\beta$ parameter, is different at $2.7\sigma$ significance between SNe~Ia with blue and red colors. The authors consider this difference insignificant and conclude that the data matches surprisingly well with a single $\beta$ across all SNe. 
The authors also find that $\beta$ can be different at $2.4\sigma$ level when SNe are split by the stellar-mass of their host galaxies, and report this trend as an indication of an evolution along the host galaxy properties. 
The differences in $\beta$ across SN color and host properties are both indications of the dust-based hypothesis discussed in \BSdust, and therefore motivates further analysis.
However, they report a notable, similarly-sized separation between host-binned SN~Ia residuals for both bluer and redder colors --- closer to the black curves in Fig.~\ref{fig:hypotheses_illustration}.  Furthermore, \GinolinColor finds no significant difference in the scatter of the Hubble residuals for blue and red SNe, which would be another key signature of the dust-based model. Additionally, their measurement of the size of this separation, often denoted as the $\gamma$ parameter, is notably larger ($\sim$0.15~mag) than previously reported values (e.g., $\gamma \sim 0.04-0.07$~mag in the DES sample; \citealt{Vincenzi24}).

The difference in the measured $\gamma$ size may be due to inconsistent definitions of $\gamma$ between the literature.
\GinolinColor introduces a technique to fit all parameters --- the stretch-luminosity relation $\alpha$, color-luminosity relation $\beta$, as well as the host-dependent correction size $\gamma$ --- simultaneously. This technique also involves fitting for parameters for each individual SN, amounting in a $3000+$ parameter fit.
Besides the difference in the approaches to avoid optimization biases, this method is different from recent studies of the same topic, such as \BSdust, \cite{Popovic21} or \cite{Popovic23}, due to the interpretation of the host-dependent correction. In \GinolinColor, the host correction $\gamma$ is part of the pre-determined functional form of standardization using $\alpha$ and $\beta$.
The latter method, in contrast, $\gamma$ is measured on final residuals after correcting the nonlinear, color-dependent trends, in addition to the Tripp standardization. 
This results in different definitions of $\gamma$ between \GinolinColor and prior works in the literature, due to the imposed functional forms in \GinolinColor and the fitting/measurement order. It is therefore important to compare values, such as $\gamma$, in a consistent manner across all samples and studies.

Still, regardless of the specific fitting methods or the measured value of a parameter, the most important element of the SN~Ia analysis is the parameterization of the model that captures the relationship between properties of the supernova, as well as properties of the host-galaxies. A successful model that fits better to the data can be characterized by a reduction of the chi-square value (or an improvement of likelihood), and comparing this reduction to the number of additional parameters can test the statistical significance of the improvement.
This improvement provides a more objective and direct comparison of models or theories than debating whether a certain measurement of a feature is significant or not.

In this paper, we re-analyze the volume-limited SNe~Ia in ZTF DR2 sample to provide the following:
(A) perform analysis in a manner consistent with analyses in the literature of other samples to enable direct comparison (Sec.~\ref{sec:analysis});
(B) study the effect of different analysis techniques to understand the cause of differences (Sec.~\ref{sec:differences}); and
(C) directly compare different models that attempt to characterize the Hubble residuals and discuss which model is preferred by the data (Sec.~\ref{sec:models}).
These items will help us interpret the studies of host-galaxy dependent models at a glance, particularly whether the ZTF~SNe indicate potentially different trends or the differences can be traced to different parameterizations.  We briefly discuss the stretch-dependence of the profile in Sec.~\ref{sec:stretch}.
In Sec.~\ref{sec:conclusion}, we summarize our findings, discuss their implications, and explore the possible future works.

We note that, in this work, we do not create full-scale simulations \citep[e.g.,][]{Kessler_Scolnic_2017_BBC} necessary for the cosmological applications, but we strongly encourage the community to create simulations of the ZTF sample, as it enables further in-depth analysis similar to \BSdust and \cite{Popovic23}.

\section{Analysis: Light Curve to the Hubble Residuals} \label{sec:analysis}

We follow the same analysis path as recent cosmology analyses with SNe Ia like that from DES \citep{Vincenzi24} and Pantheon+ \citep{Scolnic22,Brout22}.  We re-fit light curves released by ZTF with the SALT2 model from \cite{Betoule14} as implemented in the SNANA software package; this is the same model used in \GinolinColor. A SALT2 fit returns four parameters: the time of peak brightness $t_0$, the overall amplitude $x_0$, the stretch parameter $x_1$, the color parameter $c$.  The parameter $x_0$ is often translated to a brightness $m_B$.  We add typical quality cuts to the sample of $|x_1|<3$ and $|c|<0.3$.  \GinolinColor allows for an extended color range up to $c\le0.8$, but here we choose $c\le0.3$ to ensure consistency with previous analyses.  We include a redshift cut of $0.025<z_\mathrm{CMB}<0.060$ where the lower limit reduces sensitivity to peculiar velocities and the upper limit is at the expected volume limit of $z=0.06$, consistent with \GinolinColor.

To convert SALT2 parameters into distances, the simplest form is known as the Tripp equation \citep{Tripp99}
\begin{equation}\label{eq:Tripp}
\mu = m_B + \alpha x_1 - \beta \times c - M_B\ ,
\end{equation}
where $\mu$ is the distance modulus, $\alpha$ and $\beta$ are correlation-coefficients, and $M_B$ is the absolute magnitude of a SN Ia.
In recent analyses like Pantheon+, two further corrections $\gamma$ and $\Delta_B$ are added as 
\begin{equation}\label{eq:BBC}
\mu = m_B + \alpha x_1 - \beta c - \gamma p - \Delta_B - M_B\ ,
\end{equation}
where $\gamma$ is often considered a difference in SN Ia luminosity dependent on host property (e.g., mass, color), and $\Delta_B$ is a bias correction predicted by simulations for a measured value of $x_1$, $c$, the host property, and the redshift $z$ of the SN. The correction by $\gamma$ term depends on the pre-computed value $p$, a binary-like indicator of host property for each SN. This indicator splits SNe into two populations based on the host properties. In \GinolinColor, authors do not include bias corrections ($\Delta_B$), but do often fit for $\gamma$. 

There are four host properties included in ZTF DR2 used to calculate $\gamma$: the host galaxy mass, the host color, the local mass, and the local color. Of these, masses correspond to the stellar mass in the log scale $\logmass$ (dex) and colors are measured in $g-z$ (mag). The local values are measured based on an aperture near the SN site. In the following sections, we refer to these host parameters as logmass, color, local logmass, and local color.
The split points are fixed at logmass~$=10.0$, color~$=1.0$, local logmass~$=8.9$, and local color $=1.0$, which all approximately split the data into similarly sized sub-groups. 

In this work, we follow the parameterization used in \GinolinColor. This includes the following:
(1) The $p$ value that considers uncertainty in host property measurement. For instance, an SN with host logmass~$=9.8\pm0.2$ receives $p=0.84$ since the split point logmass~$=10$ is at its 84th percentile in uncertainty, instead of $1.0$.
(2) No bias corrections $\Delta_B$. While bias correction is often important to cosmology, it is sufficient to omit it for this volume-limited sample as the purpose of this work is to test the consistency of the models found in previous analyses and check if they would still apply to the ZTF sample. 
(3) The uncertainty in the distance modulus. Similarly to Pantheon+ and many other analyses, this uncertainty depends on the coefficients $\alpha$ and $\beta$, as well as the intrinsic scatter size $\sigma_\text{int}$:
\begin{equation}\label{eq:sigma_mu}
    \sigma_\mu^2 = 
    \begin{pmatrix}
        1 & \alpha & -\beta
    \end{pmatrix}
    \mathbf{C}
    \begin{pmatrix}
        1\\\alpha\\-\beta
    \end{pmatrix} + \gamma^2\sigma_p^2 + \sigma_\text{int}^2 
\end{equation}
where $\mathbf{C}$ is the covariance matrix between $m_B$, $x_1$, and $c$. Following \GinolinColor, we assume the uncertainty in the binary indicator of host properties $\sigma_p$ is very small and thus negligible. The additional uncertainty determined during bias correction $\sigma_\Delta$ is not applicable for \GinolinColor since no bias correction is performed. Following \GinolinStretch, Eq.~\ref{eq:sigma_mu} does not include redshift uncertainty.
In this work, we fix the intrinsic scatter at $\sigma_\text{int}=0.17$, which makes the reduced chi-square of the residuals close to unity in the later analyses. As we discuss in Sec.~\ref{sec:differences_binning} and Sec.~\ref{sec:models}, we fix this parameter-dependent uncertainty at $\alpha=0.15$ and $\beta=3.00$ and use it for model comparisons and Monte Carlo bootstrapping only. Other analyses use statistics of median values that does not depend on the uncertainty, so these choices made to be consistent with \GinolinColor and \GinolinStretch have negligible effect on our conclusions.

\begin{figure*}
    \centering
    \includegraphics[width=\linewidth]{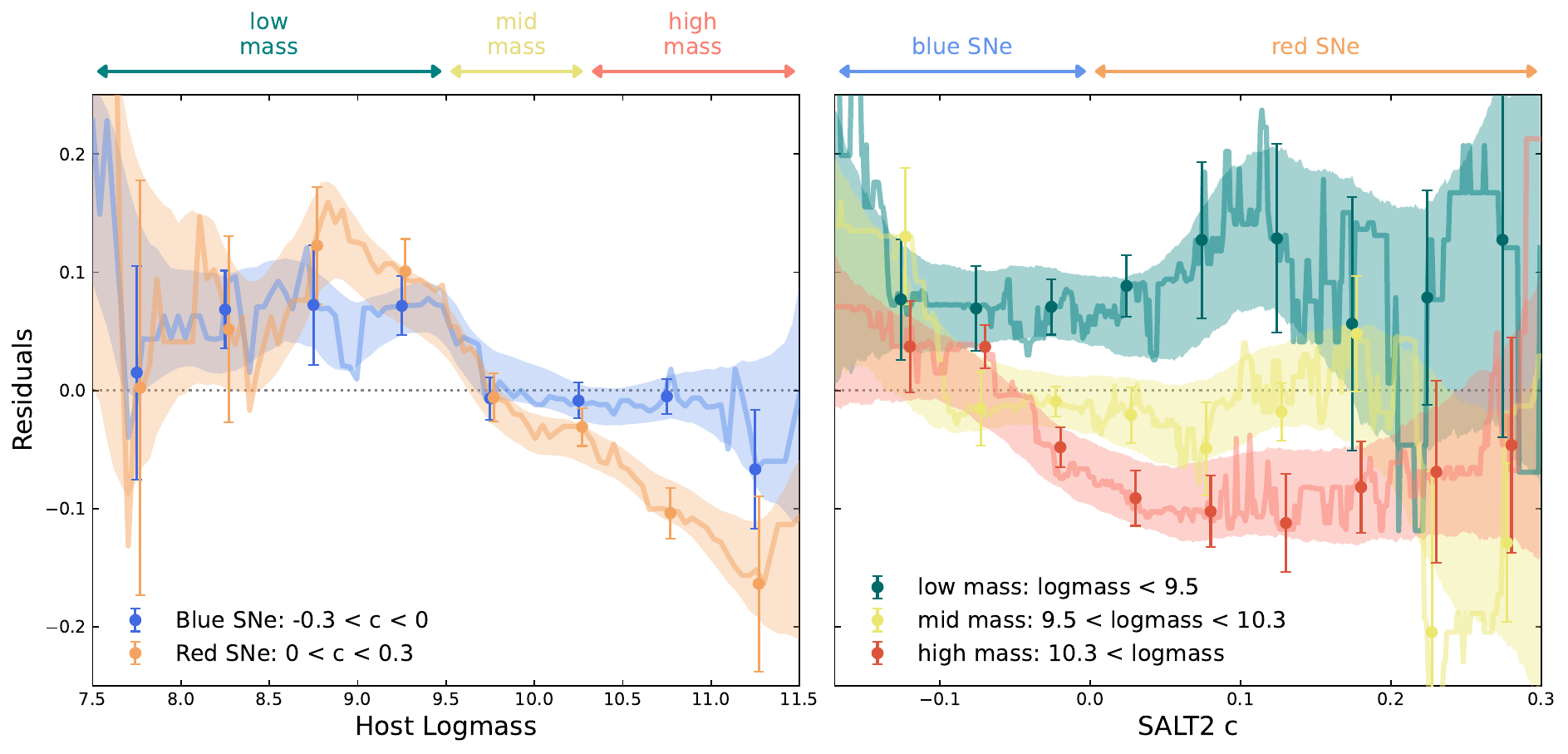}
    \caption{(Left) Hubble residuals as a function of host logmass, when split for blue and red colors at $c=0$.  We only fit $\alpha$ and $\beta$ to compute the residuals plotted: no mass-step ($\gamma$ term) is added. We note there is no obvious `step' for the red SNe.  (Right) A further split of the plot shown in panel 1a, where now low mass (logmass~$<9.5$, medium mass ($9.5\le\text{logmass}<10.3$), and high mass ($10.3\le$~logmass) are all shown.  This is consistent with the panel to the left that there is a smooth dependence on mass for the Hubble residuals of red SNe, and a consistent convergence for Hubble residuals of blue SNe.}
    \label{fig:alligator_mass}
\end{figure*}

To ensure consistency with previous literature, we employ the optimization technique used in the SALT2mu (\citealp{Marriner11}; see \citealp{Scolnic14} for summary) analysis, which does not depend on a particular choice of cosmological models.

In the following sections, we analyze and discuss the Hubble residual, the residual of the computed distance moduli from the $\lcdm$ model at the observed redshift,
\begin{equation} \label{eq:HR}
    \text{HR}=\mu_\text{obs} - \mu_{\lcdm}(z_\text{CMB})
\end{equation}
Additionally, we calculate the median redshift dependency of the computed residuals
\begin{equation} \label{eq:HR_corr}
    \text{HR}_\text{corr} = \text{HR} - \Delta_\mu(z_\text{CMB})\ .
\end{equation}
This redshift dependency is computed from the residuals of best-fit Tripp standardization (Eq.~\ref{eq:Tripp}) and fixed throughout the subsequent analyses (i.e., not re-computed for each correction model to avoid affecting the comparison). The value of $\Delta_\mu(z)$ is mostly small and flat across redshifts, except an offset towards lower redshift, $\Delta_\mu(z\lesssim0.04)\approx0.07$.

\begin{figure*}
    \centering
    \includegraphics[width=\linewidth]{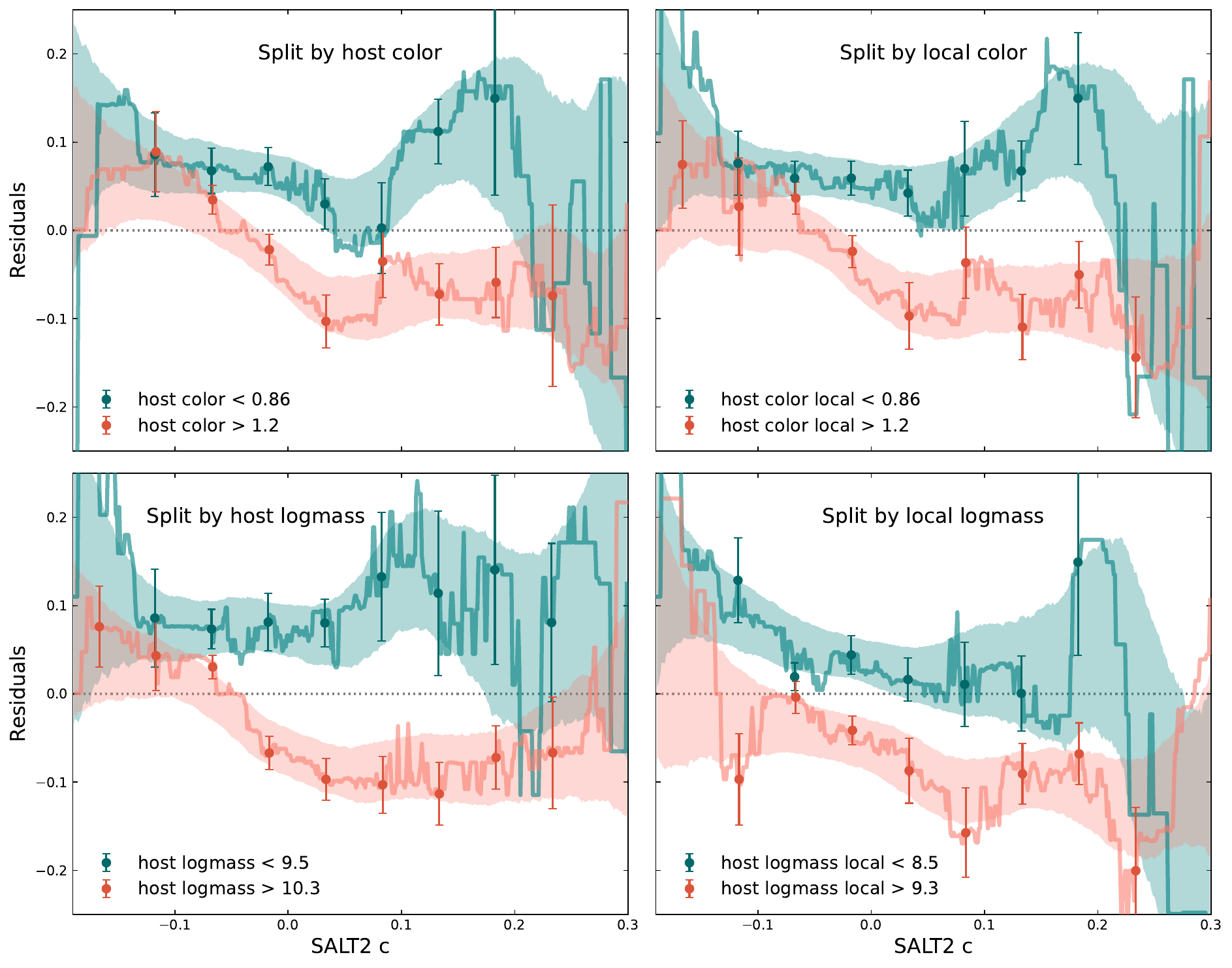}
    \caption{Color-dependent residuals, split by ``step'' locations of host property tracers. }
    \label{fig:alligators_all}
\end{figure*}

\subsection{Dependance of Hubble Residuals Offsets on Supernova Colors and Host-Galaxy Properties} \label{sec:result_alligator}

We present the Hubble residuals $\text{HR}_\text{corr}$ as a function of color and host logmass in Fig.~\ref{fig:alligator_mass}.
The shown profiles are calculated by placing bins at each location and taking the median value. We also show a continuous profile obtained by moving a fixed-size bin over small increments (rolling median) in addition to a set of literature-like standard bins. We discuss the binning techniques in Sec.~\ref{sec:differences_binning}.

The trends visible on the left panel reveal a few key points --- first, we indeed find the host-mass dependency of the residuals. However, the ``step'' location does not seem to be centered at $\log(\frac{M^*}{M_\odot})=10$ but rather near $\log(\frac{M^*}{M_\odot})\approx9.6-9.7$. Second, there is a significant slope past $\log(\frac{M^*}{M_\odot})\approx9.5$ for redder SNe. The change in the residual is as large as the ``step'' size beyond the split point, indicating that the mass-dependency is not as simple as a single step \citep[e.g., see ``mass-slope'' model:][]{Uddin_2020_massslope}. This slope-like feature is less evident for blue SNe. 

The right panel of Fig.~\ref{fig:alligator_mass} reveals an even more important feature --- the separation between $\logmass<9.5$ and $\logmass>10.3$ groups (see the right panel, Fig.~\ref{fig:alligator_mass}) show a clear color-dependency. We see a convergence near $c\approx-0.1$ -- the residuals in SNe have negligible host-dependency for blue ($c\approx-0.1$) SNe ($\Delta \text{HR} = 0.05\pm0.06\,$mag; also see discussions in Sec.~\ref{sec:differences_gamma}), while there is a large host dependency for red ($c\approx0.1$) SNe. The quick divergence past $c>-0.1$ brings the separation between these two groups to $\Delta \text{HR} = 0.23\pm0.07\,$mag level, a $0.18\pm0.09$ mag increase in separation compared to the blue ($c\approx-0.1$) side.
This trend has been called ``alligator mouths'' due to the convergence of Hubble residuals at $c\sim-0.1$ with a gradual divergence toward redder color.  
We observe color-dependent variations of residual profiles with other tracers as shown in Fig.~\ref{fig:alligators_all}, though the size of host-dependency varies between different tracers.
These findings are consistent with \cite{BroutScolnic21}, and they suggest the use of a more complex model that can accommodate the color-dependency. In Sec.~\ref{sec:differences}, we discuss the effects of using more complex model and its implication.

\subsection{Dependence of Hubble Residuals Scatter on Supernova Colors and Host-Galaxy Properties} \label{sec:result_scatter}

\begin{figure*}
    \centering
    \includegraphics[width=2\columnwidth]{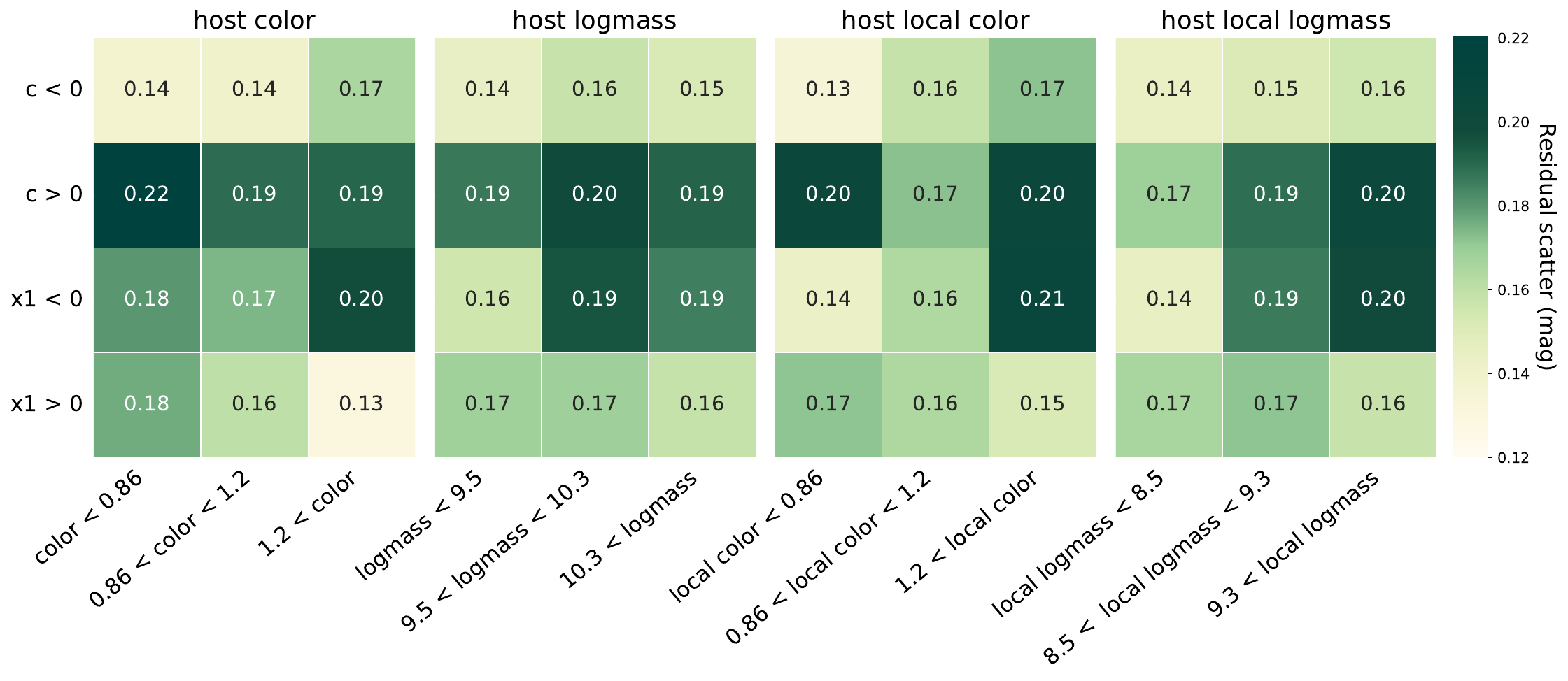}
    \caption{The scatter of residuals in subsamples, split by their light curve parameters ($c$, $x1$) and host properties. The scatter size is measured with nMAD, $1.48\cdot\,\text{Med}(|\text{HR}|)$. Residuals (HR) are calculated with Model 1 in Table~\ref{tab:model_comparison}. We bootstrap the residual values within each bin to reduce the effect of small number statistics. Note the number of data points vary between bins, and the size of difference may not translate to the significance of difference. See Sec.~\ref{sec:result_scatter} for discussions.}
    \label{fig:scatter}
\end{figure*}

In addition to quantifying residual trends through binning, the scatter size can highlight the complexity of the residuals and its dependency on color and host properties.
We measure the scatter size using normalized median absolute deviation ($\text{nMAD}=1.48\cdot\text{Med}\left(\left|\text{HR}\right|\right)$). In Fig.~\ref{fig:scatter}, we show the measured scatter sizes of subsamples, split by their light curve parameters (color and stretch) and host properties. The residuals are calculated with the best-fit $\alpha$ and $\beta$ values for the simplest Tripp correction ($\alpha=0.141,\ \beta=3.006$; discussed in Sec.~\ref{sec:models}), and each subset is further median-subtracted (to separate the discussion from Sec.~\ref{sec:result_alligator}).
Due to the redshift range of our SNe samples ($0.025\leq z \leq 0.060,\ \bar{z}\approx0.045$), the scatter size is sensitive to uncertainty in redshift measurements. To reduce the effect of incorrect redshift estimates, we further limit the samples to SNe with small uncertainties in redshifts, $\sigma_z < 0.001$, which passes 458 out of 796 SNe.
In addition, we bootstrap the residual values $\text{HR}_\text{corr}$ within each bin to reduce the effect of small number statistics in the measurement of nMAD values.

The measured scatter sizes in Fig.~\ref{fig:scatter} indicates that blue SNe ($c<0$, $\sigma \sim 0.15$) and red SNe ($c>0$, $\sigma \sim 0.20$) have a clear difference in the scatter size. The Brown–Forsythe Test \citep{levene1960,Brown_Forsythe_1974} measures the difference in scatter size between each population; within each host property bin, this test confirms the difference in scatter size at $2.0\sigma$ (p-value$\approx0.045$) significance on average. In all cases, redder SNe ($c>0$) have larger scatter. When all host property bins are combined (i.e., simple split by $c$ only), the significance increases to $3.4\sigma$ ($p=0.00055$).
This difference is likely not due to the larger measurement error: the distribution of the distance modulus uncertainty $\sigma_\mu$ between $c<0$ and $c>0$ subsamples are indistinguishable according to the KS-test \citep{Massey_1951_KStest}, which measures the difference between two distributions as a whole, as its significance is $0.76\sigma$ ($p=0.45$). The same KS-test for color error $\sigma_c$ alone also yields the same observation ($0.83\sigma$, $p=0.40$).

The observed trend is different when SNe are split by $x_1$. When all host properties are combined (i.e., simple $x1$ split), the Brown–Forsythe test measures no significant difference between $x_1<0$ and $x_1 > 0$ subsamples ($1.09\sigma$, $p=0.28$). Instead, the difference between scatter size seems to become larger as the host property evolves (redder, more massive); in the rightmost bins in each host property types, the scatter is consistently larger in $x_1<0$ subsamples compared to $x_1>0$. However, it is unclear how much of this is due to small sample statistics --- the number of SNe with $x_1>0$ are significantly skewed towards younger, less evolved host types in this dataset, and the limited data in bins like $x_1>0$ and $\text{color}>1.2$ makes the separation not as significant as it seems. For the most evolved host property bins (rightmost bins of each group in Fig.~\ref{fig:scatter}), the significance of difference between $x_1<0$ and $x_1>0$ are $1.5\sigma$ ($p=0.12$) for color, $0.65\sigma$ ($p=0.51$) for logmass, $1.75\sigma$ ($p=0.07$) for local color, $1.31\sigma$ ($p=0.18$) for local logmass, all lower than the $2\sigma$ average difference in $c$--splits.

\section{Tracing the Differences} \label{sec:differences}

Our findings, that there is a significant color-dependency in mass-steps and other similar splits by other host properties, is different from some of the conclusions presented in \GinolinColor.
In this section, we aim to understand the effects of different analysis techniques and study how they can lead to different conclusions.
We trace the driving causes of the differences to binning methods, split locations, fitting technique, and cuts. Below we describe each item, discuss the difference, and propose improved methods and the benefits they offer.

\begin{figure*}
    \centering
    \includegraphics[width=\linewidth]{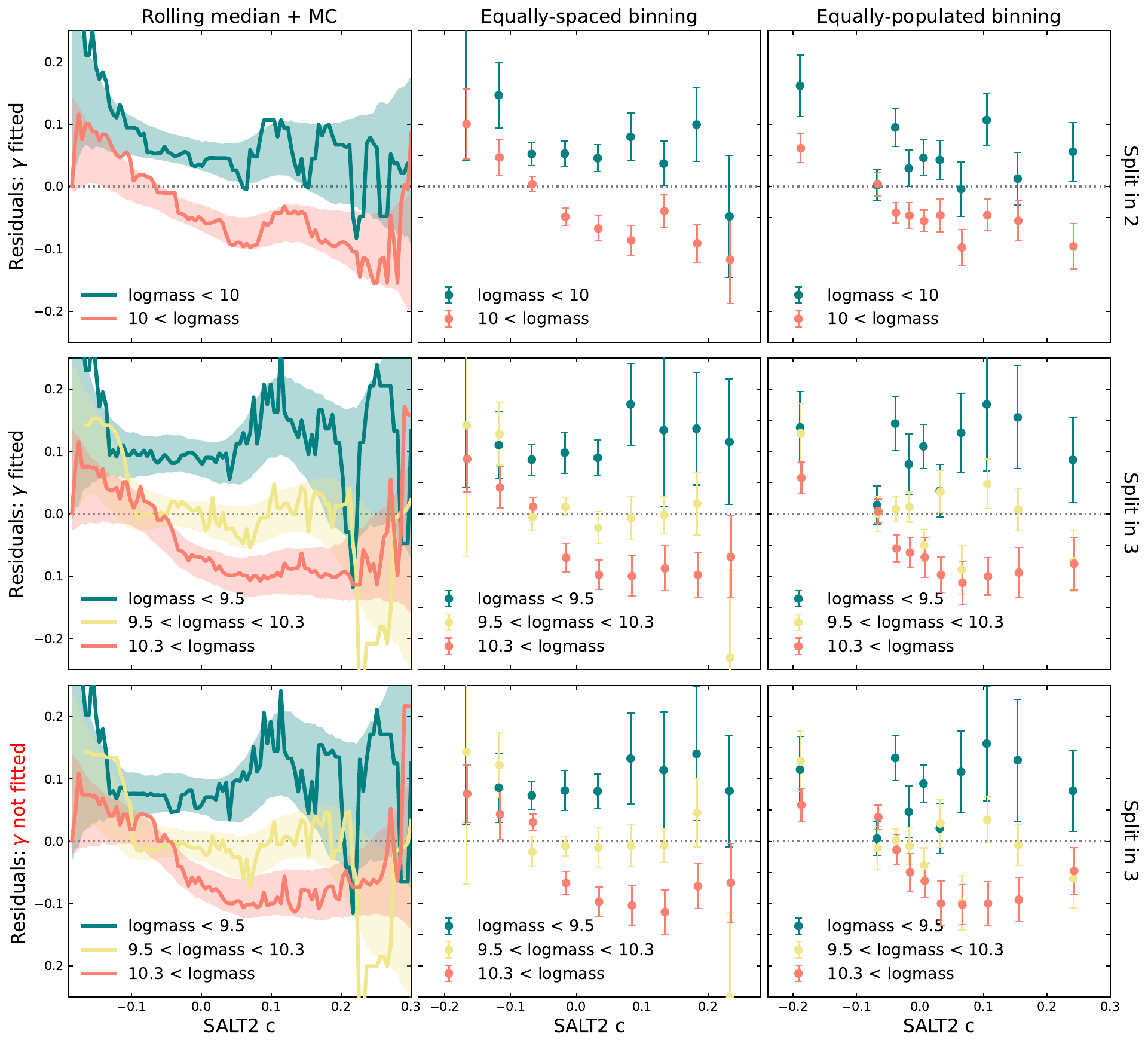}
    \caption{Comparison of fitting and binning methods that highlight the causes of different interpretations in \GinolinColor. In all panels, we plot the Hubble residual against SALT2 color, where the residual does \textit{not} contain the $\gamma$ term (i.e., before mass-step correction). Column-wide comparison highlights the dependence on ``measured'' profiles on binning methods, and row-wide comparison demonstrates the effects of splits and fitting methods. \textit{Right column:} equally-populated binning, similar to \GinolinColor. \textit{Middle column:} equally-spaced binning, similar to \BSdust. \textit{Left column:} equally-spaced rolling binning, with the shaded region obtained through Monte Carlo bootstrapping. \textit{Top row:} single split point at $\logmass=10$, similar to \GinolinColor. The parameters $\alpha,\beta,\gamma$ are all fitted simultaneously (but $\gamma$ is not applied), similarly to \GinolinColor. \textit{Middle row:} Same fit as the top row but split in 3, treating the intermediate population separately. \textit{Bottom row:} Same spilt as the middle row but with two-parameter fit (no $\gamma$ fitted). The difference between middle and bottom row is also shown in Fig.~\ref{fig:residual_separation}.}
    \label{fig:binning-comparison}
\end{figure*}

\begin{figure}
    \centering
    \includegraphics[width=\linewidth]{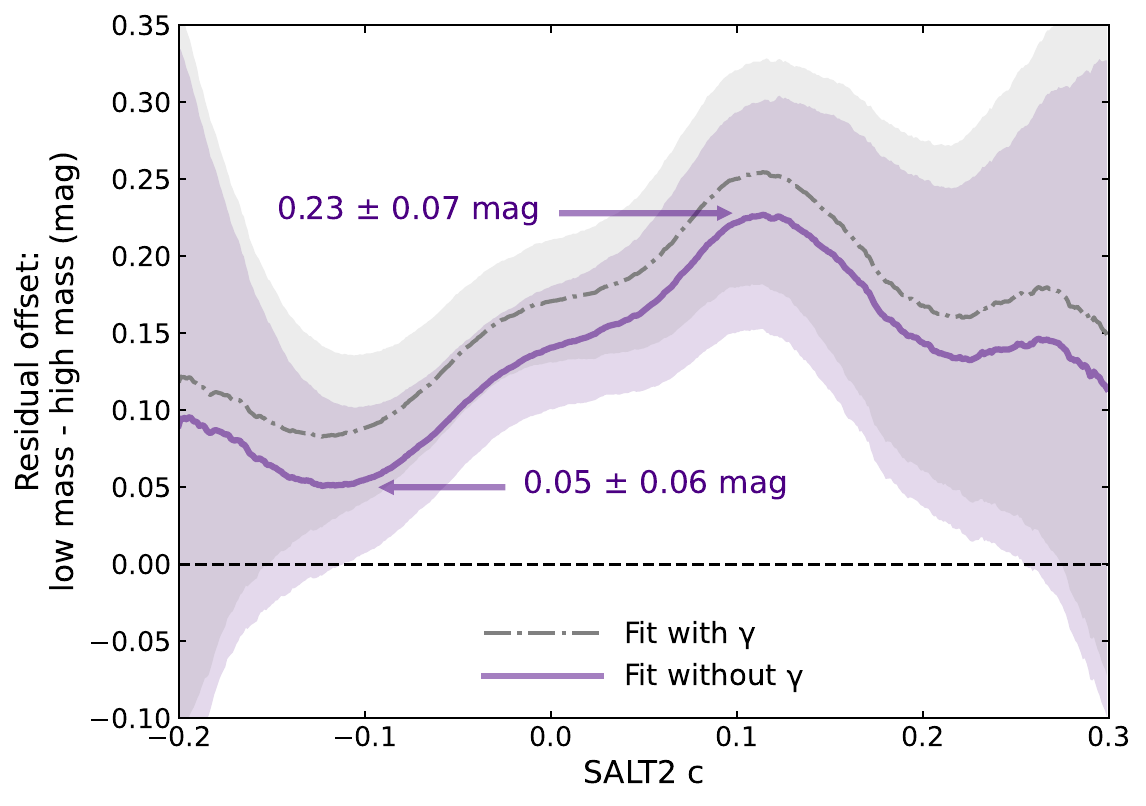}
    \caption{The separation of the Hubble residuals between low-mass (logmass $<$ 9.5; green curves in Fig.~\ref{fig:binning-comparison}) and high-mass (10.3 $<$ logmass; red curves in Fig.~\ref{fig:binning-comparison}) populations, shown as a function of color. Dotted and solid curves represent the residuals calculated with and without $\gamma$ parameter, respectively (see Sec.~\ref{sec:differences_gamma} for discussion). The significant difference in the separation sizes, as highlighted in the figure, necessitate a color-dependent model instead of a mass-step, which attempt to model this profile with a single horizontal line. See Sec.~\ref{sec:differences_binning} for binning method used to obtain the profiles. }
    \label{fig:residual_separation}
\end{figure}

\subsection{Binning Methods}\label{sec:differences_binning}
In Fig.~\ref{fig:binning-comparison}, we present the impact of binning on the measured profiles. In the middle and right panels, we compare the ``alligator'' profiles measured over equally-spaced and equally-populated bins. For equally-spaced bins, we divide SNe into a grid of bins with width of $0.05$-mag, and for equally-populated bins, we divide SNe into groups of 10th, 20th, ..., 100th percentiles. Since there are less SNe towards $c\approx-0.2$ or $c\approx0.3$, there is a large difference in binning spacing between these two methods. 
In the left panels, we show the same quantities calculated by the rolling median method \citep[see][for review]{Tukey_1977_ExploratoryDataAnalysis,Huang_1979_MedianFilter},  which moves a fixed-sized (0.05 mag) bin along the color axis and calculate the median value at each location. This method visualizes the sensitivity of the binned quantities on the location, and when combined with the Monte Carlo bootstrapping we describe later, it cleanly reveals the underlying trend in data (shaded regions).

The measured profiles show noticeable differences. This is expected, because 
(1) equally-spaced binning on unevenly sampled data statistically underestimates the measured trend, and 
(2) data is highly sensitive to the binning locations, regardless of the binning size.
The rolling-bin plot on the left panels reveal the sensitivity of binning locations.
Due to the decreasing population of SNe towards the tails and increasing scatter size toward redder color (see Sec.~\ref{sec:result_scatter} for discussion on scatter size), the standard binning methods may not exhibit the larger underlying trends, depending on where exactly the bin falls onto, and it makes the comparison between literature difficult.

To reduce this effect and reveal the larger trends, we further add the shaded regions in the left panels that represent the 1-$\sigma$ regions of the rolling-median profile obtained through Monte Carlo (MC) re-sampling of the data, which acts as a data-driven smoothing of the profile. For this task, we randomly draw color and the residual from their observed value and uncertainty, obtain the profile, and repeat this process for $\sim1000$ times to obtain the mean profile and its standard deviation. This technique significantly reduces the statistical noise coming from the discrete nature of binning, and it allows us to distinguish statistical fluctuations from the true trends. The uncertainty of the residual for MC include the observational uncertainty only (i.e., $\sigma_\text{MC} = \sqrt{\sigma_\mu^2 - \sigma_\text{int}^2}$), since intrinsic scatter size depends on the host- (and color-) dependent correction. 

\subsection{Host Split Locations}\label{sec:differences_split}
Another major contributing factor to the difference from \GinolinColor is the location(s) of the ``step'' at which host properties are split into two groups. In Fig.~\ref{fig:alligator_mass}, it is evident that the residual drop is not clearly located at $\log(\frac{M^*}{M_\odot})\approx10$, which has been the conventional split point in the past decades \citep[e.g.,][]{Kelly2010}. As we discuss later, this is driven by the mismatch of mass-step model from the true underlying profile.
If the host mass--residual profile is best described by a model other than the mass-step model, comparing the two populations split at $\logmass \approx10$ can misguide the analysis, since each group contain the intermediate populations that, on average, does not show the true trend unique to each group.
We therefore split SNe into three groups -- low mass ($\log(\frac{M^*}{M_\odot})\le9.5$), intermediate mass ($9.5<\log(\frac{M^*}{M_\odot})\le10.3$), and high mass ($10.3<\log(\frac{M^*}{M_\odot})$). The lower split point is chosen to approximate the onset point at which the residual starts to drop down (left panel, Fig.~\ref{fig:alligator_mass}), and the upper split point is chosen to approximate the point at which the significance divergence between blue and red SNe occur. The number of SNe in each group are 225, 276, and 295, respectively.
As shown in the middle rows in Fig.~\ref{fig:binning-comparison}, these new splitting locations clear up the ``alligator mouth'' pattern across all three binning methods. The intermediate populations now lie between the large divergence between low-mass and high-mass populations, which further confirms that the underlying trend is not a mass-step.

\subsection{$\gamma$: different technique, different definitions} \label{sec:differences_gamma}
In the discussions above, we noted that the step models do not best-describe the underlying profiles our data exhibit. To evaluate the effect of fitting the step parameter $\gamma$ simultaneously with $\alpha$ and $\beta$, we compare two sets of residuals (middle and bottom rows, Fig.~\ref{fig:binning-comparison}) computed with combinations of ($\alpha$, $\beta$) fitted with and without $\gamma$ (see Sec.~\ref{sec:models} for comparisons of different models). Note the $\gamma$-correction is \textit{not} applied for the residuals in Fig.~\ref{fig:binning-comparison}.

The best-fit values of $\alpha$ change when we fit $\gamma$ simultaneously, which is in contrast to the $\beta$ values that do not seem to be affected by the presence of $\gamma$. The comparison of the offsets between high-mass and low-mass populations are shown in Fig.~\ref{fig:residual_separation}.
While the effect is smaller than the other causes discussed in the previous subsections, we observe a noticeable difference in the overall separation size when $\gamma$ is fitted simultaneously: the minimum separation between low-mass and high-mass populations is $\sim 0.05$ mag when $\gamma$ is not fitted, and the value increases by $\sim60$\% when $\gamma$ is fitted simultaneously.
We also note that Fig.~\ref{fig:residual_separation} also reveals the significance of color dependency, with the peak separation near $c\approx0.1$ at $0.23\pm0.07$ mag. 
The mass-step model attempts to model this curve with a single horizontal line, and it explains the larger value reported in \GinolinColor compared to the previous results. In dust-based models like \BSdust, the color-dependency is removed first, which nearly flattens this curve at the minimum level (i.e. $\sim 0.05$ mag) before measuring the leftover mass-step size. We discuss the fitting methods in detail in Sec.~\ref{sec:models}.

\section{Analytic models for residuals}\label{sec:models}

\begin{deluxetable*}{cl|c|r|rrrrrrrrrrr}
    \tabletypesize{\footnotesize}
    \tablewidth{0pt}
    \tablecaption{Comparison of different models and their performances.\label{tab:model_comparison}}
    
    \tablehead{
    \colhead{\#} & \colhead{Model} & \colhead{Indicator} & \colhead{Cut} & 
    \colhead{$\alpha$}& \colhead{$\beta$} &\colhead{$\gamma$}&
    \colhead{nMAD} & \colhead{$\sum\chi^2$} & \colhead{N$_\text{param}$} & \colhead{$\Delta\chi^2_{1\rightarrow i}$}& \colhead{$\Delta\chi^2_{\text{step}\rightarrow i}$}}
    
    \startdata 
    0 & No correction               & - & - & - & - & - & 0.375 & 4312.2 & 0 & - \\
    1 & 2-param Tripp               & - & - & 0.141 & 3.006 & - & 0.185 & 901.1 & 2 & 58.3$^{a}\sigma$\\ 
    \hline
    2 & Tripp + step (\GinolinColor)& local color & 1.0 & 0.161& 3.050& 0.143& 0.172& 837.1& 2+2 & 7.6$\sigma$ & -\\
    3 & Tripp + step (fit)          & local color & 1.0 & 0.173& 3.061& 0.153& 0.174& 839.2 & 2+2 & 7.5$\sigma$ & -\\
    4 & Tripp + BS21-lite           & local color & 1.0 & 0.173& 3.061& (0.079)$^{b}$& 0.165 & 815.1 & 2+8 & 7.8$\sigma$ & 3.3$\sigma$\\
    5 & Tripp + Host2D            & local color & -   & 0.141& 3.006 & - & 0.174& 818.2 & 2+11 & 7.2$\sigma$ & 2.2$\sigma$\\
    \hline
    6 & Tripp + step                & global color & 1.0 & 0.174 & 3.062 & 0.147 & 0.177 & 836.9 & 2+2 & 7.6$\sigma$ & -\\
    7 & Tripp + BS21-lite           & global color & 1.0 & 0.174 & 3.062 & (0.081)$^{b}$ & 0.177 & 817.0 & 2+8 & 7.7$\sigma$ & 2.8$\sigma$\\
    8 & Tripp + Host2D            & global color & -   & 0.141 & 3.006 & - & 0.171 & 825.8& 2+11 & 6.7$\sigma$ & 0.4$\sigma$\\
    \hline
    9 & Tripp + step                & local mass & 8.9 & 0.157 & 3.145  & 0.108 & 0.184 & 875.2 & 2+2 & 4.6$\sigma$ & -\\
    10 & Tripp + BS21-lite           & local mass & 8.9 & 0.157 & 3.145  & (0.054)$^{b}$ & 0.174 & 846.1 & 2+8 & 5.8$\sigma$ & 3.9$\sigma$\\
    11 & Tripp + Host2D            & local mass & -   & 0.141 & 3.006 & - & 0.171& 821.2 & 2+11 & 7.0$\sigma$ & 5.5$\sigma$\\
    \hline
    12 & Tripp + step                & global mass & 10 & 0.165 & 3.022 & 0.127 & 0.180 & 841.7 & 2+2 & 7.3$\sigma$ & -\\  
    13 & Tripp + BS21-lite          & global mass & 10 & 0.165 & 3.022 & (0.073)$^{b}$ & 0.168 & 823.7 & 2+8 & 7.3$\sigma$ & 2.5$\sigma$ \\
    14 & \textbf{Tripp + Host2D}  & global mass & -  & 0.141 & 3.006 & - & 0.173 & \textbf{804.9} & 2+11 & \textbf{7.9$\mathbf{\sigma}$} & \textbf{4.0}$\mathbf{\sigma}$\\
    15 & Tripp + step + Host2D    & global mass & 10 & 0.165 & 3.022 & 0.127 & 0.170 & 807.6 & 2+2+11 & 7.5$\sigma$ & 3.4$\sigma$\\  
    \enddata
    \tablecomments{The best-fit parameters, residuals, and chi-square values for 796 SNe~Ia. $\chi^2$ are calculated with a fixed uncertainty calculated at $\alpha=0.150$, $\beta=3.00$, and $\sigma_\text{int}=0.170$. The last two columns $(\Delta\chi^2)$ show the sigma-significance of improvement each model achieves compared to the baseline model $\#1$ and the step model, respectively. $^{a}$For model \#1 only, the shown value is a comparison against the no-correction case (model \#0). $^{b}$ These numbers represent the separation between low-mass and high-mass populations after the color dependency is corrected, but this term is included in the polynomial functions and is not a fitted parameter.}
\end{deluxetable*}

\begin{figure*}
    \centering
    \includegraphics[width=\linewidth]{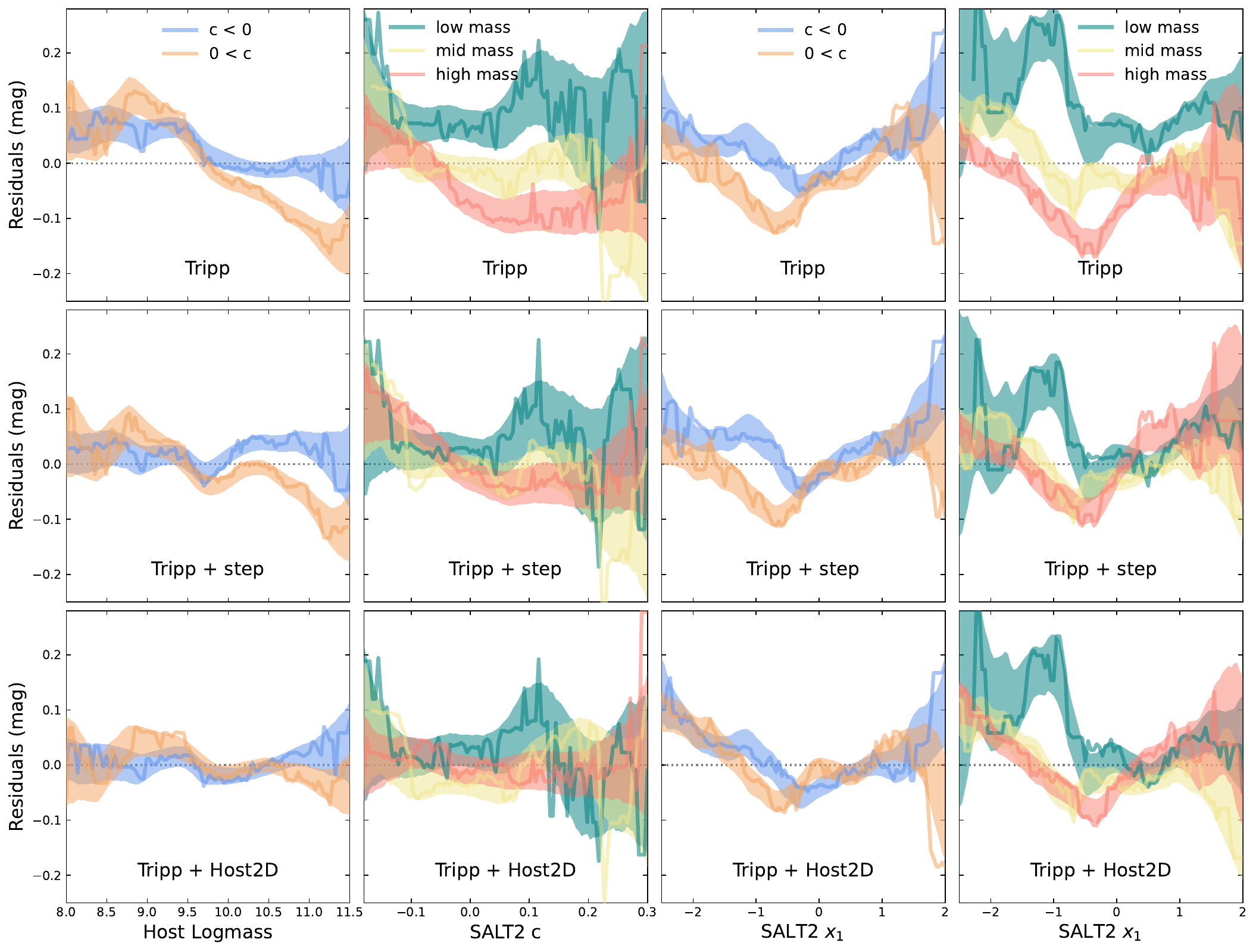}
    \caption{Comparison of correction methods and remaining residuals over host logmass, SN~Ia color, and SN~Ia stretch ($x_1$) values. \textit{Top row:} the residuals after applying the best-fit Tripp standardization (model \#1). \textit{Middle row:} the residuals after applying the mass-step model (model \#12). \textit{Bottom row:} the residuals after applying the Host2D model (model \#14).}
    \label{fig:tripp_step_BBC}
\end{figure*}

The goal of SN~Ia standardization is to account for known variation of luminosities based on observable parameters. With the complexity in data we find in Sec.~\ref{sec:result_alligator}, it is evident that corrections dependent on host properties are necessary. In this section we compare existing models: the basic Tripp standardization, the step models discussed in \GinolinColor and \GinolinStretch, and a color-dependent step model by \BSdust. In addition, we present Host2D, an expansion of \BSdust model that accounts for the continuous change over a range of host properties. 

We present the best-fit parameters for each model, optimized using the SALT2mu method (as described in Sec.~\ref{sec:analysis}), in Table~\ref{tab:model_comparison}, and visualize the comparison of selected models in Fig.~\ref{fig:tripp_step_BBC}. We evaluate the performance of each model by comparing the total chi-square value $\sum_\text{SN}\chi^2$ of the residuals after subtracting the model, with the chi-square for each SN calculated as 
\begin{equation}\label{eq:chisq}
    \chi^2 = \left[\frac{m_B + \alpha x_1 - \beta c - \gamma p - M_B - \mu_\text{cosmo}(z)}{\sigma_\mu(\alpha=0.15,\,\beta=3.0)}\right]^2\ ,
\end{equation}
where $\sigma_\mu$ (Eq.~\ref{eq:sigma_mu}) is fixed at $\alpha=0.15$ and $\beta=3.0$ so that $\chi^2$ values provide a pure comparison of residuals. This uncertainty includes an arbitrary $\sigma_\text{int}=0.170$ mag intrinsic scatter.
When a model offers an improved $\chi^2$ value, we evaluate its significance considering the increased number of parameters (split location and step size). 
The significance is measured by the $p$-value\footnote{
When computing $p$-values or $\sigma$-values in regimes of high significance, the associated probabilities become exponentially small. Consequently, numerical instabilities may arise unless appropriate computational strategies and precision are employed. We use \texttt{mpmath} package to ensure sufficient numerical precision is maintained throughout our computations.} may , which measures the probability of such improvement occurring by chance
\begin{equation}
    p=P(\chi^2>\Delta\chi^2; \Delta \nu)\ ,
\end{equation}
where $\Delta \chi^2$ is the change in chi-square, $\Delta \nu$ is difference in number of parameters, and $P$ measures the upper-tail cumulative probability of the chi-square distribution with $\Delta \nu$ degrees of freedom. For a more intuitive discussion, we also provide the value of $\sigma$--significance, which we calculate from the $p$-value.
In Table~\ref{tab:model_comparison}, we include this $\sigma$-improvement of each model in comparison to the baseline model and the `step' models.
The following sections describe each model and discuss the improvements they offer. 

\subsection{Base model: Tripp standardization}
Tripp standardization (Eq.~\ref{eq:Tripp}) is based on empirical findings that SN~Ia color and stretch correlate with luminosity. This model accounts for the majority ($\gg50$\%) of the intrinsic variation between SNe~Ia. 
The limitation of this model lies in its simple approximation that stretch-luminosity and color-luminosity is linear.
The best-fit parameters for this model, $\alpha=0.141$ and $\beta=3.006$, produces $\sum\chi^2=901.1$ --- a $\sim80$\% in rediction of $\chi^2$ with two parameters, which is by far the most efficient improvement at $58.3\sigma$ significance.
We use this model as a basis of comparison against other models. 

\subsection{Step models}
Step models introduce the host galaxy properties as an additional observable (Eq.~\ref{eq:BBC} without $\Delta_B$). Typically the step locations are determined separately or fixed at commonly used values. For our analysis, we use the locations same as \GinolinColor and \GinolinStretch.

We first test the best-fit values of $\alpha$, $\beta$, $\gamma$ parameters reported in \GinolinColor (model 2), using local color as a host property tracer. This indeed reduces the scatter ($\sum\chi^2=837.1$) compared to the base model at $7.6\sigma$ significance.
The best-fit parameters for the same model determined by SALT2mu method (model 3) yields a similar result ($\sum\chi^2=839.2$, $7.5\sigma$ improvement). 
The step model with different tracers perform at a similar level, with global color at $7.6\sigma$ and global mass at $7.3\sigma$ improvement, except for the local mass that significantly underperforms below $5\sigma$.

This result confirms the reports in \GinolinColor that the step model most efficiently reduces the residual scatter among the four host property tracers used.

\subsection{BS21-lite: Color-dependent mass-step}
\BSdust proposed an improved version of the mass-step model, which introduces the alligator-mouth shape --- the color-dependency of the separation between low-mass ($\logmass<10$) and high-mass ($\logmass>10$) populations. 
While simulations of dust models and comparison with the ZTF data is beyond the scope of this work, we showcase the impact of color-dependent model by fitting a simplified \BSdust--like model.
Note this model, as well as any model discussed in \GinolinColor only accounts for correcting the median residual at each color and host property, while the full \BSdust model constructs a covariance matrix based on observed light curve parameters and host logmass, which further reduces the effective size of scatter.
We split SNe by their host tracer, similarly to step models, to fit two third-order polynomial curves that model color-dependent residuals in each. 
We employ the parameterization similar to Eq.~\ref{eq:BBC},
\begin{equation}\label{eq:BS21}
    \mu = m_B + \alpha x_1 - \beta c - \boldsymbol{\tilde\gamma}(c,p) - M_B\ ,
\end{equation}
where $\boldsymbol{\tilde\gamma}(c,p)$ is a function that chooses between two polynomials $P_\text{low}$ and $P_\text{high}$ representing the color dependency of the split populations (e.g., $\logmass<10$ and $\logmass>10$):
\begin{equation}
    \boldsymbol{\tilde\gamma}(c,p) = p\cdot P_\text{low}(c) + (1-p) \cdot P_\text{high}(c)\ .
\end{equation}
This model achieves the largest improvement when local color is used as a host tracer, with $\sum\chi^2=815.1$, which is a 7.8$\sigma$ improvement from the Tripp model and 3.3$\sigma$ improvement from the local color-step model.
Across all host tracers, this model achieves significant improvements over the simple step models ($2.5-3.9\sigma$), which confirms the importance of treating the color dependency. Additionally, once the color dependencies are removed, the step size $\gamma$, which is the minimum separation between $P_\text{low}$ and $P_\text{high}$, is significantly smaller than the values measured by the step model, in agreement with \BSdust and other previous studies of this model.

\subsection{Host2D: A Continuous 2D Map over\\*Host Property and SN Color}
\begin{figure*}
    \centering
    \includegraphics[width=\linewidth]{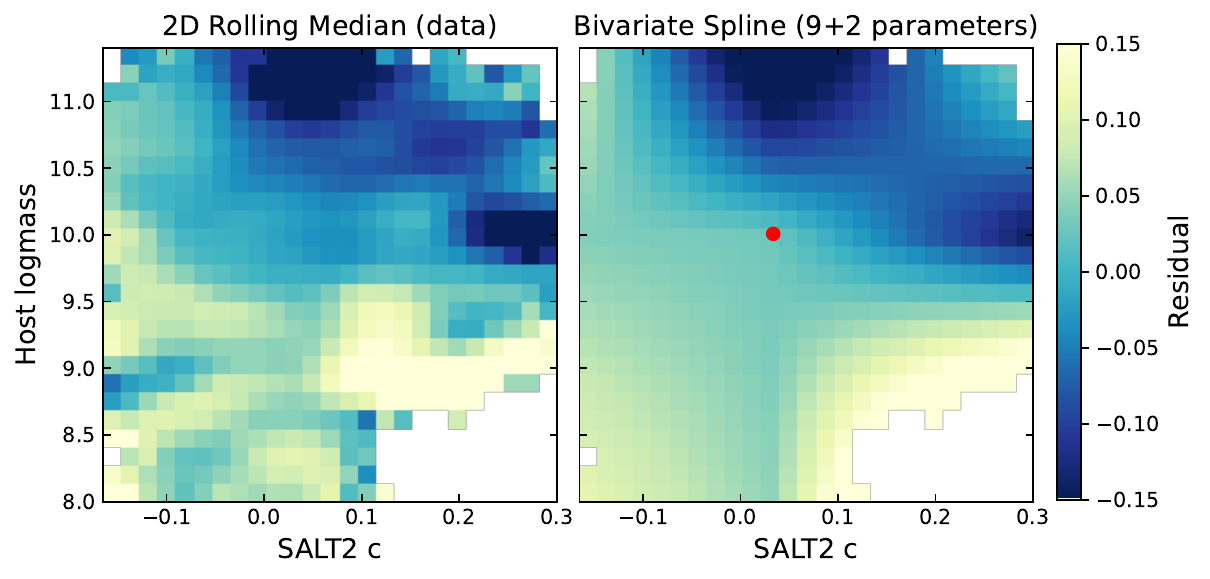}
    \caption{An example of 2D map created for Host2D. \textit{Left:} the data map, calculated with a rolling-bin + MC as discussed in Sec.~\ref{sec:differences_binning}. Bin locations with less than 3 SNe are excluded and marked as white. \textit{Right:} the fitted Host2D map. The red dot represents the location of the spline knot that minimizes the $\chi^2$. The map is comprised of linear functions connected at the knot.}
    \label{fig:Host2D}
\end{figure*}

\begin{figure*}
    \centering
    \includegraphics[width=\linewidth]{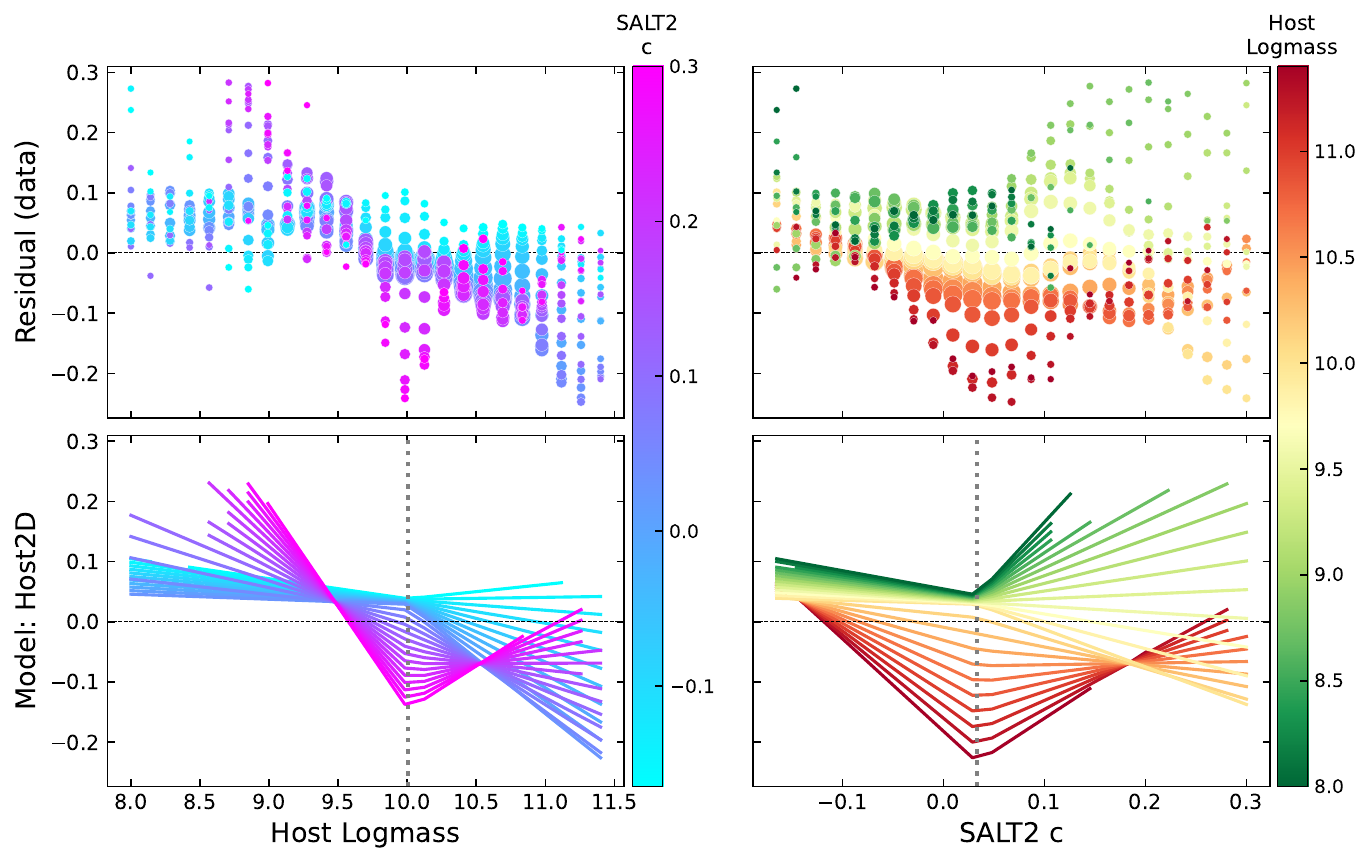}
    \caption{A different visualization of Fig.~\ref{fig:Host2D}, in the style similar to Fig.~\ref{fig:alligator_mass}. \textit{Top:} Measured residuals in 2D rolling bin (left panel of Fig.~\ref{fig:Host2D}; each dot corresponds to each grid point). Dot size represents the number of SNe in each bin. \textit{Bottom:} color-logmass surface modeled by Host2D (right panel of Fig.~\ref{fig:Host2D}). Vertical line represents the knot location, at which two bilinear functions, left and right side, are connected.}
    \label{fig:Host2D_trace}
\end{figure*}
In Sec.~\ref{sec:result_alligator} and Sec.~\ref{sec:differences_split}, we discussed the significance of intermediate population, which indicates the continuity of host property-dependent residuals, rather than a simple separation in a step-like manner. We test the impact of a continuous correction as an expansion of \BSdust model by empirically modeling the continuous surface. 
We implement Host2D, a correction of the Hubble residuals based on a low-parameter representation of the two-dimensional map of the Hubble residuals over SN~Ia color and host tracer. 
This is similar to the analysis in \cite{Wiseman22} which did not utilize simulations for the volume-complete Dark Energy Survey supernova sample, as there are no selection effects. 

To construct the Host2D model, we first create a 2D map of residuals using the methods described in Sec.~\ref{sec:differences_binning} (rolling median and Monte Carlo). The residuals are calculated with the best-fit 2-parameter Tripp model ($\alpha=0.141, \beta=3.006$). We restrict the data map to be comprised of bins with more than three SNe to avoid overfitting to the sparsely populated regions. We fit a bivariate spline \citep[see, e.g.,][for review]{deBoor_1978_spline,Dierckx_1981_spline} with a single knot and polynomial order of 1 (i.e., piecewise bilinear function) to model the 2D map. Analytically, this is written as a function of SN color $c$ and host property value $h$,
\begin{equation}
    \bar{\gamma}(c,h) = \sum_{i,j=0}^{2}\mathbf{a}_{ij}B_i(c,c_0)B_j(h,h_0)\ ,
\end{equation}
where $\mathbf{a}_{ij}$ is a $3\times3$ coefficient matrix, and the spline basis functions $B_0(x,x_0)=1,\ B_1(x,x_0)=x,\ B_2(x,x_0)=\max\{x-x_0,0\}$ determine the location of the given coordinate $x$ with respect to the knot location of $x_0$.
This model has 11 free parameters -- 3$\times$3 for the coefficient $\mathbf{a}$ and 2 for the knot locations along color- and host property-axes ($c_0$, $h_0$). We use \texttt{LSQSphereBivariateSpline} implementation in the \texttt{Scipy} package and a custom routine\footnote{The code is publicly available at \url{https://github.com/SterlingYM/Host2D}} to optimize all 11 parameters.
The domain of the best-fit map is strictly limited to match the data map to avoid extrapolation. 

An example of the Host2D map is shown in Fig.~\ref{fig:Host2D} for the case of using host logmass as a tracer. Additionally, an alternative, prehaps more intuitive visualization of the Host2D map, as well as the data map, is shown in Fig.~\ref{fig:Host2D_trace}. This plot provides the slice in both host logmass and SALT2 color directions, similarly to Fig.~\ref{fig:alligator_mass}.
Subtracting this model from the residuals (using host logmass; Model \#14), the chi-square value of the corrected data reduce to $\chi^2=804.9$, which is a $7.9\sigma$, 4.0$\sigma$, and 3.4$\sigma$ improvement from the baseline model (\#1), mass-step model (\#12), and the BS21-lite model (\#13), respectively. 
The best-fit location of the spline knot (the coordinate at which separate bilinear functions are connected) is determined to be $c_0=0.04, \ h_0=10.03$ -- a location that matches the onset points where the divergence pattern between blue/red or low-mass/high-mass populations change, as shown in Fig.~\ref{fig:alligator_mass} (which is more vividly illustrated in Fig.~\ref{fig:Host2D_trace}).

\subsection{Comparison and the Implication}
The test results demonstrate that a color-dependent, continuous model using host logmass (Host2D; Model \#14) offers the best, most significant reduction of scatter size at $7.9\sigma$, $4.0\sigma$ improvement from the baseline and the mass-step model, respectively.  This model also outperforms the color-dependent, step-like \BSdust model by $3.4\sigma$, indicating that treating the continuity between low-mass and high-mass populations provides a significant benefit.

The Host2D model \textit{does not} require $\gamma$, and in fact, applying $\gamma$ correction before Host2D worsens the result, both in $\chi^2$ and especially the significance of reduced scatter size (model \#15). This is likely because adding a discontinuity to the data by applying the step function to a naturally continuous data reduces the modeling power of the Host2D method.

For a volume-limited dataset which is relatively free from observational biases, the Host2D model that accounts for the color-dependent, continuous features is shown to outperform other previously proposed models. 
This result also underscores the difference between the exploratory analysis, such as binned statistics in Sec.~\ref{sec:analysis}, and an analytical modeling with the goal of cosmological applications.
Further studies need to be conducted to integrate this Host2D model with BBC before it can be used for larger, full range of surveys that require bias corrections. Until then, the \BSdust model offers a well-tested, superior alternative, which takes advantage of the color-dependent step sizes as a $2.5-3.9\sigma$ improvement to the simple, single-step models.
If one simply wants to \textit{approximate} the mean offsets between subsamples binned by host properties for an exploratory, illustrative purpose, it is important to define what the reported value of the ``step'' size represents, as the value depends on the fitting method. Finally, if there is any significant step that occurs in a scale much more acutely than what Host2D is capable of capturing, we recommend the step function to be subtracted from the residuals as a final step after all \BSdust- or Host2D-like corrections are applied, though the data we analyzed in this work does not show evidence for such step after other corrections.

\subsection{The Stretch Dependency:\\*toward the Host3D model} \label{sec:stretch}
In \GinolinStretch, authors suggested that the broken-alpha law, a Tripp--like standardization with the slope $\alpha$ changing at a certain location of $x_1$, may fit better to the data. While it is out of scope for this work, we confirm a large, V-shape residuals in Fig.~\ref{fig:tripp_step_BBC} over $x_1$ axes (right two columns).

Most notably, after applying the Host2D model with logmass (model \#14), the high-mass ($\logmass>10.3$) subsample exhibits a significant, linear trend between $x_1\approx-2.5$ and $x_1\approx0$  with a total of $0.183\pm0.045$ mag, a $>4\sigma$ level of stretch-dependency. This trend, however, is not shared across all subsamples split by host properties: the low-mass ($\logmass<9.5$) population exhibits a significantly different profile with a sudden drop rather than a slope, with a $>3.5\sigma$ separation of $\Delta\text{HR} = 0.18\pm0.05$~mag near $x_1\approx-1$.
When SNe are split by color, bluer SNe exhibits a much less significant but nonzero separation from redder SNe at a similar location of $x1\approx-1$ at $\sim2\sigma$ significance.
The difference in the residuals provides three observations: 1. there is a large stretch-dependency in the residuals that cannot be removed by any of the models we tested in this work. 2. we can leverage the stretch dependency to further improve the modeling of residuals. 3. the stretch-dependency is not uniform across all host types (and possibly color), indicating that a multi-dimensional model beyond a simple modification of the stretch-luminosity relation will provide the maximum improvement.
We consider this as a hint and a motivation towards the added treatment over the $x_1$ space, especially in a multi-dimensional manner similar to the Host2D, and we encourage the community to investigate further.

\section{Conclusions} \label{sec:conclusion}

We re-fit the ZTF DR2 SN~Ia light curves to study the color-dependency of host--residual profiles, such as mass step. Our dataset is similar to \GinolinColor, with SNe~Ia within $0.025 < z < 0.060$ (lower bound from the public release) and whose host galaxy properties (color, local color, mass, and local mass) are included in the ZTF DR1 dataset. 
Below we summarize our findings and lookout to further improvements.

\textbf{1. The Trend in Data.} When the residuals are split by their SALT2 color or host tracer, we find significant host- and color-dependent profiles that differ from the simple step-like model. In particular, when the SNe are split into two groups by host logmass ($\logmass < 9.5$ and $\logmass>10.3$), we find a significant color dependency, with a small ($\Delta=0.05\pm0.06$ mag) separation between low-mass and high-mass groups for bluer SNe near $c\approx-0.1$ and a large $0.23\pm0.07$ mag separation between for redder SNe near $c\approx0.1$. For all host groups, the scatter size is different between red and blue SNe, with redder SNe having larger scatter by $\sim30\%$ at $3.4\sigma$ level.
This trend, which is consistent with findings in \BSdust, suggests that we need a correction model that can accurately portray multi-dimensional, nonlinear features in residuals that depend both on color and the host tracer. Additionally, we identify a significant intermediate populations between these two groups, which suggests that the host dependency is a gradual, continuous effect, and an improved analysis technique may be needed to fully capture these trends.

\textbf{2. Differences.} We attribute the differences of the conclusions in \GinolinColor from ours to the differences in analysis technique, not the characteristics of data. We demonstrate that a set of improved, robust statistical methods and analysis techniques uncover the features and trends in data that simpler analyses may overlook:

\textit{Binning:} The measured color-dependent profiles are sensitive to the binning locations, and a single set of bins may not accurately depict the underlying trend. This issue cannot be mitigated by equally-populated binning method, as the uneven distribution of data biases the measured trend, and the measurement is still sensitive to the exact bin locations.
We propose a workaround to this issue by using rolling median with Monte Carlo sampling, which allows one to reduce the discrete nature of binning and obtain a statistically more reliable trend.

\textit{Split Locations:} The measured behavior of intermediate populations (e.g., $9.5 < \logmass < 10.3$) suggests a gradual change in the residual (see Fig.~\ref{fig:alligator_mass}), and including them in low- or high-mass populations (or other equivalent host splits) misleads the analysis to a significantly suppressed separation and color dependency. Splitting the SNe into three groups and treating the intermediate population separately significantly improves the visibility of the divergence between low-mass and high-mass populations.

\textit{Fitting Method:} Fitting the step parameter $\gamma$ simultaneously with the standard Tripp parameters $\alpha$ and $\beta$ impacts the best-fit $\alpha$ parameter, which leads to a difference in the measured color-dependent profile. Including $\gamma$ in the fit enlarges the separation between low-mass and high-mass populations. This can be caused by the mismatch of the imposed mass-step model to the data, which is consistent with our findings above. 

\textbf{3. Models.} We compare 15 standardization and residual-correction models to test their ability to efficiently capture the trend in data and reduce the residuals, including Tripp standardization, step models, \BSdust, and Host2D --- an improved, continuous \BSdust--like model over the color--host property space.
Our results show that added complexity generally improves the ability to capture the trends without losing the efficiency, with Host2D (with logmass as a tracer) achieving the largest reduction in $\chi^2$ at the highest efficiency ($7.9\sigma$), followed by \BSdust and step models. 
Of all host tracers tested, step model prefers local color as a tracer as reported in \GinolinColor, while Host2D and \BSdust yields the best result with host logmass as a tracer.
The $>4\sigma$ improvement of $\chi^2$ from a mass-step model to the Host2D model highlights the importance of treating the color-dependency and continuous trend, rather than a uniform, single step. While Host2D does not require $\gamma$, a care should be taken when one wishes to compare the size of $\gamma$, as the definition of step size and $\gamma$ changes depending on which model is used and whether the color-dependency is corrected before the measurement or not.

\textbf{4. Lookout.} The discussions in this work highlights the significance of the choice of analysis methods on the interpretation of data and its implication on the models we select. With improved technique, we identify the color-dependency and intermediate populations; accounting for both of them, as shown with the Host2D model, achieves a significant reduction in the residual size. 
Our work underscores the difference between visualization and modeling; while the binning methods provide useful insights into the trend in data, it does not indicate that the functional form of the model should follow the binned (i.e., step) pattern.
The model should be chosen by its fit to data, with the number of parameters in consideration.

Additionally, we identify the stretch dependency at $>3\sigma$, similar to \GinolinStretch, hinting that such modeling methods can be further expanded to larger parameter spaces, and there is still a large room for improvements of SN~Ia standardization.
We encourage the community to utilize robust statistical tools and analysis methods presented here, as well as the proposed continuous model. Such additions and improvements, especially when combined with full-scale simulations, will significantly enhance the exploration of the large volume-limited SNe~Ia samples like ZTF DR2 and provide most useful comparisons of scatter models.

\begin{acknowledgements}
We thank Madeleine Ginolin for comments and discussions on SN samples, fitting methods, and parameterization of models. We thank Dillon Brout for discussions on the BS21 model. We thank Adam Riess for helpful comments and continued support and guidance. Y.S.M. is grateful to Siyang Li, Louise Breuval, Javier Manniti, Wenlong Yuan, and Ruoxi Wang for their continuous support.
D.S.~is supported by Department of Energy grant DE-SC0010007, the David and Lucile Packard Foundation, the Templeton Foundation, and Sloan Foundation. 
\end{acknowledgements}

\bibliographystyle{aasjournal}
\bibliography{main}{}

\end{document}